\documentclass[pra,twocolumn,showpacs,floatfix]{revtex4}
\usepackage{graphicx}
\begin{document}

\title{Rotational properties of non-dipolar and dipolar 
Bose-Einstein condensates confined in annular potentials}
\author{E. \"O. Karabulut$^{1,4}$, F. Malet$^2$, G. M. Kavoulakis$^3$, and S. M. Reimann$^1$}
\affiliation{$^1$Mathematical Physics, Lund University, LTH,
P.O. Box 118, SE-22100 Lund, Sweden \\
$^2$Department of Theoretical Chemistry and Amsterdam Center for Multiscale Modeling, 
FEW, Vrije Universiteit, De Boelelaan 1083, 1081HV Amsterdam, The Netherlands \\
$^3$Technological Educational Institute of Crete, P.O. Box 1939, 
GR-71004, Heraklion, Greece \\
$^4$Physics Department, Faculty of Science, Selcuk University,
TR-42075, Konya, Turkey}
\date{\today}

\begin{abstract}
We investigate the rotational response of both non-dipolar 
and dipolar Bose-Einstein condensates confined in an annular 
potential. For the non-dipolar case we identify certain
critical rotational frequencies associated with the formation 
of vortices. For the dipolar case, assuming that the dipoles 
are aligned along some arbitrary and tunable direction, we 
study the same problem as a function of the orientation angle 
of the dipole moment of the atoms.
\end{abstract}

\maketitle

\section{Introduction}

In recent experiments on ultra-cold atomic gases the
confinement of atoms in toroidal traps has become possible
\cite{rtr}. In such confining potentials persistent currents 
have also been created and observed \cite{Ryu, BPh, Foot, ZH},
in close analogy to semiconductor nanostructures \cite{Vief, 
Mann}. The simplicity of these systems makes them ideal for 
studying effects associated with superfluidity, including 
rotational properties, persistent flow, etc. 

In another series of recent experiments it has become 
possible to trap atoms \cite{Grie} and molecules \cite{JYe} 
with a nonzero dipole moment. Contrary to most of the 
previously realized experiments, in which the particles
had zero dipole moment and the usual contact interaction 
modeled the effective atom-atom/molecule-molecule interaction 
rather well, in the dipolar case the interaction potential 
is long-ranged and anisotropic. As a consequence of these
differences, dipolar gases present new properties that
allow for the study of novel and interesting effects. 

In three dimensions the head-to-tail alignment of the dipoles 
causes instabilities towards collapse \cite{Lah}, while 
low-dimensional confinement tends to make the system more 
stable \cite{Brunn}. Quasi-two- or quasi-one-dimensional 
confinement thus suppresses the instabilities associated
with the dipolar interaction and is advantageous in that 
respect. 

Recently, several properties of dipolar gases have been 
examined in effectively two-dimensional systems. Among these, 
the rotational properties of dipolar degenerate gases have 
received considerable attention. In particular, the rotational 
properties of dipolar Bose-Einstein condensates have been 
investigated in different confining geometries, including 
axially-symmetric harmonic \cite{Aba3}, elliptic harmonic 
\cite{FMG11}, and toroidal traps \cite{Aba10, MKR, AB1, AB2, 
AB3}. Also, recently it has become possible to create a 
Bose-Einstein condensate of dysprosium 164, with a magnetic  
moment of roughly three times larger than chromium 52, 
which increases the dipole coupling constant by roughly a 
factor of ten \cite{BLev}. 

Dipolar Bose-Einstein condensates confined in toroidal 
traps have been investigated both within the mean-field 
approximation \cite{Aba10}, and in the few-particle limit 
\cite{Zollner}. These systems have been shown to exhibit 
some very interesting effects due to the interplay between 
the nontrivial topology of the ring-like confinement and 
the anisotropic nature of the dipole-dipole interaction. 
More specifically, when the dipole moments have a non-zero 
in-plane component, the effects associated with breaking 
the axial symmetry of the system have been shown to have 
potential applications in, e.g., Josephson-type 
oscillations and self-trapping phenomena \cite{AB1, AB2, 
Xio}.

Motivated by the experimental progress mentioned above 
on toroidal traps and on dipolar gases, we study below 
the rotational properties of a Bose-Einstein condensate 
that is confined in an annular-like potential, first 
considering a condensate that consists of non-dipolar 
atoms \cite{Smerzi, Pedri, Woo} and then addressing the 
dipolar case. Our calculations are closely-related 
with those of Ref.\,\cite{Fet67}, which analysed the 
lowest-energy state of a uniform (non-dipolar) rotating 
superfluid confined in a hard-wall annulus. In this 
study it was shown that as the rotational frequency 
increases initially there are vortices that are located 
in the region of zero density. As the rotational 
frequency increases, vortices start to form in the 
region of nonzero density. With increasing angular 
frequency of the trap the real vortices eventually 
form a circular array, or even multiple arrays of 
vortices. These theoretical predictions were in good 
agreement with the experimental results on liquid Helium 
that followed afterwards \cite{Don67}. In what follows 
below we refer to the vortices that are located in the 
region of negligible density as ``phantom" vortices and 
the ones that are located in the region of non-negligible 
density as ``real" vortices. 

The paper is organized as follows. In Sec.\,II. we present
our model. In Sec.\,III we investigate the rotational 
response of a non-dipolar gas. In the spirit of 
Ref.\,\cite{Fet67} we identify the corresponding critical 
rotational frequencies. In Sec.\,IV we investigate the effect
of the dipolar interaction, assuming that the dipole moment of 
the atoms is oriented along some arbitrary direction due to
an external polarizing field. We demonstrate that the vortex
structures depend strongly on the orientation angle of the 
dipole moment of the atoms. Finally in Sec.\,V we summarize 
our results. 

\section{Model}

The annular potential that we consider is modeled via a 
combination of harmonic potentials in the transverse direction 
and in the direction perpendicular to the plane of motion 
of the atoms,
\begin{eqnarray}
V({\bf r}) = V_r({\bf r}_{\perp}) + V_z(z) =
\frac 1 2 M \omega_0^2 (r_{\perp} - R)^2 
+ \frac 1 2 M \omega_z^2 z^2.
\nonumber \\
\end{eqnarray}
Here $z$ is the symmetry axis of the trap, which is 
also the axis of rotation of the annulus. In addition 
${\bf r}_{\perp}$ is the position vector on the $x$-$y$ 
plane, while $\omega_0$ and $\omega_z$ are the trap 
frequencies. The ratio $\omega_z/\omega_0$ is chosen 
to be equal to 100, which makes the motion of atoms 
quasi-two-dimensional, since $\hbar \omega_z$ is the 
largest energy scale in the problem. The ratio $R/a_0$
is chosen equal to 4, where $a_0 = \sqrt{\hbar/(M \omega_0)}$
is the oscillator length that corresponds to the frequency 
$\omega_0$ and mass $M$. Figure 1 shows schematically the 
corresponding annular-like potential and also the external
field that is assumed to polarize the (dipolar) atoms. Without 
loss of generality, this field is taken to be on the $xz$-plane, 
forming an angle $\Theta$ with the $x$-axis.

\begin{figure}[t]
\centerline{\includegraphics[width=6cm,clip]{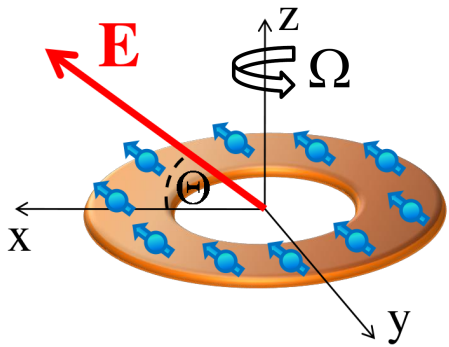}}
\caption{(Color online)
Schematic illustration of the rotating quasi-two-dimensional 
annular trap. The atoms move on the $x$-$y$ plane with an 
angular frequency $\Omega$, which is along the $z$ axis, in 
an annulus of mean radius $R$ and a width that is set by the 
oscillator length $a_0 = \sqrt{\hbar/(M \omega_0)}$. In 
the case of dipolar atoms, their dipole moment is oriented 
on the $x$-$z$ plane, forming an angle $\Theta$ with the $x$ 
axis due to an external (magnetic or electric) field ${\bf E}$.} 
\label{fig1}
\end{figure}

Concerning the interactions, these are modeled via a 
short-ranged contact potential -- which takes care of 
the short-range correlations -- and the usual dipole-dipole 
potential, which describes the long-range part of the 
interaction. Because of the assumed strong confinement 
along the $z$-axis, one may safely assume that the order 
parameter has a product form 
\begin{eqnarray}
  \Psi({\bf r}) = \Phi({\bf r}_{\perp}) \times \phi(z),
\end{eqnarray}
where $\phi(z)$ is the Gaussian ground state of the harmonic 
potential $V_z(z)$. This assumption allows us to reduce the 
problem from three dimensions to two dimensions for the order 
parameter $\Phi({\bf r}_{\perp})$, which satisfies the 
Gross-Pitaevskii-like equation
\begin{eqnarray}
\biggl[- \frac {\hbar^2 \nabla_{\bot}^2}{2 M} + 
V_r({\bf r}_{\bot}) 
+ V_{\rm dip}({\bf r}_{\bot}) 
+ g |{\Phi({\bf r}_{\bot})}|^2 \biggl]  
\Phi({\bf r}_{\bot}) = 
\nonumber \\ = \mu \Phi({\bf r}_{\bot}).
\label{gpe}
\end{eqnarray}
In the above equation $g = (\hbar^2/M)\sqrt{8 \pi} N 
a/a_z$, where $N$ is the atom number, $a$ is the 
scattering length for zero-energy elastic atom-atom 
collisions, $a_z$ is the oscillator length in the $z$ 
direction, $\mu$ is the chemical potential, and
\begin{equation}
V_{\rm dip}({\bf r}_{\bot}) = \int 
V_{\rm eff}({\bf r}_{\bot}-{\bf r}_{\bot}^{\prime}) \,
\left|\Phi({\bf r}_{\bot}^{\prime})\right|^2\,
d{\bf r}_{\bot}^{\prime}
\end{equation}
is the effective dipolar interaction potential, where
$V_{\rm eff}({\bf r}_{\bot}-{\bf r}_{\bot}^{\prime})$ 
is given in Ref.\,\cite{Cre10}. 

The problem thus reduces to solving the nonlocal and 
nonlinear integro-differential Eq.\,(\ref{gpe}). We 
solve it with use of a fourth-order split-step Fourier 
method within an imaginary-time propagation approach 
\cite{Chi05}. Within this method one starts with some 
initial state, which then propagates in imaginary time 
until numerical convergence is achieved. 

\section{Non-dipolar atoms in a rotating annular trap}

We first consider the case of non-dipolar atoms and 
vary the rotational frequency $\Omega$ for a fixed
interaction strength, choosing $M g/\hbar^2 = \sqrt{8 
\pi} N a/a_z = 150$. Since the ratio between the 
interaction energy and $\hbar \omega_0$ is roughly 
$N a a_0/(a_z R) \sim 10$, and also $\omega_z/\omega_0 
= 100$, there is a clear hierarchy of energy scales, 
with $\hbar \omega_z$ being roughly ten times as large 
as the interaction energy, which in turn is roughly ten
times as large as $\hbar \omega_0$. Therefore, while 
the cloud is in the ground state of the harmonic 
oscillator along the $z$ direction, it is closer to 
the Thomas-Fermi limit in the direction along the 
plane of motion.

Using the numerical procedure outlined above, the 
energetically favourable state of the cloud is 
determined in the rotating frame for a certain value 
of $\Omega$. As in Ref.\,\cite{Fet67} we identify 
three critical rotational frequencies $\Omega_{c1}$, 
$\Omega_{c2}$, and $\Omega_{c3}$; see Fig.\,2. The 
first critical frequency, $\Omega_{c1}$, corresponds 
to the formation of a ``phantom" vortex state, which 
is located at the center of the annulus. These ``phantom" 
vortices are effectively states of quantized circulation  
with integer quantum number. In this case the density 
of the cloud remains unchanged and the only change takes 
place in the phase of the order parameter, which acquires 
a jump of $2 \pi$. 

The second critical frequency, $\Omega_{c2}$, is 
associated with the formation of the first ``real" 
vortex. According to our results, this vortex state 
actually penetrates the cloud from the inner circle. 
The point where the vortex enters the cloud is arbitrary, 
as the symmetry of the Hamiltonian is broken spontaneously. 
Eventually more vortices penetrate the cloud and finally 
at the third critical frequency $\Omega_{c3}$ an array 
of vortices forms. We identify $\Omega_{c3}$ as the 
frequency for which the vortices are located (roughly) 
at the middle of the annulus. For the parameters given 
above $\Omega_{c1}/\omega_0$, $\Omega_{c2}/\omega_0$, and 
$\Omega_{c3}/\omega_0$ are found to be $\approx 0.036$, 
0.53, and 0.63, respectively. 

A direct comparison of the above numbers with those 
of Ref.\,\cite{Fet67} is not possible since in that 
case the annular potential was modeled as a hard-wall 
potential, while in our study we have assumed 
harmonic confinement and as a result the density varies 
in the transverse direction. According to \cite{Fet67}, 
neglecting logarithmic factors of order unity, $\Omega_{c1}
/\omega_0$ is $\approx a_0^2/(2 R d)$, where $d$ is the width 
of the annulus. Assuming that $d = a_0$, then $\Omega_{c1}
/\omega_0 = 1/8$, which is roughly a factor of 3.5 larger 
than the value found numerically. Also, according to the 
same study $\Omega_{c2}/\omega_0 = 2 (a_0/d)$, which is 
equal to two, and this differs by a factor of roughly 
four from the value found above. On the other hand, the 
ratio $\Omega_{c1}/\Omega_{c2} = a_0/(4R) = 1/16$, which 
is rather close to the ratio $0.036/0.53 \approx 1/14.7$. 

\begin{figure}[t]
\centerline{\includegraphics[height=6cm,width=9cm]{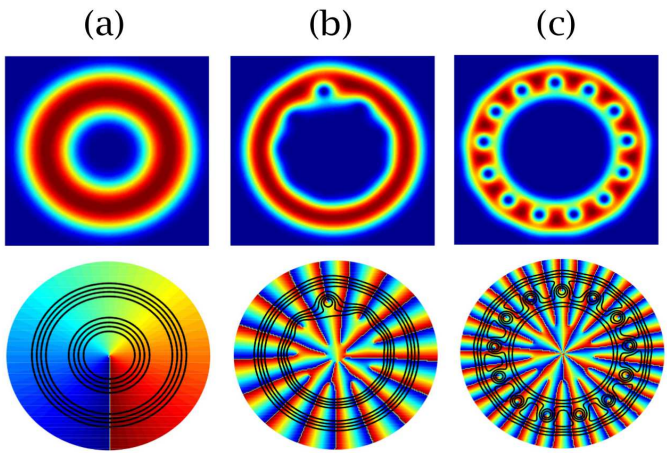}}
\caption{(Color online)
Density and phase of the order parameter corresponding 
to the three critical frequencies for the non-dipolar 
case ($D=0$) and $\sqrt{8 \pi} N a/a_z = 150$. For (a) 
$\Omega/\omega_0 = \Omega_{c1}/\omega_0 \approx 0.036$ a 
``phantom" vortex appears, for (b) $\Omega/\omega_0 = 
\Omega_{c2}/\omega_0 \approx 0.53$ the first ``real" vortex 
enters the cloud, and for (c) $\Omega/\omega_0 = \Omega_{c3}/
\omega_0 \approx 0.63$ an array of vortices forms along the 
annulus.}
\label{fig2}
\end{figure}

\begin{figure}[t]
\centerline{\includegraphics[height=13cm,width=6cm]{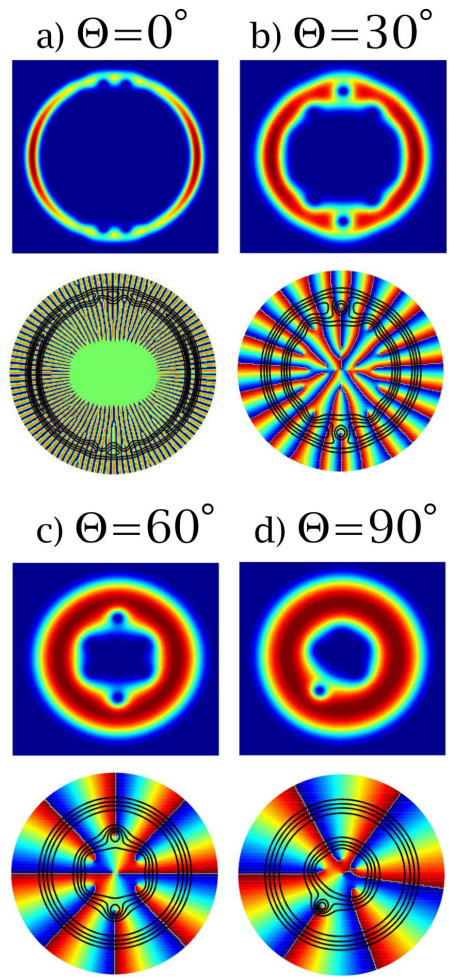}}
\caption{(Color online)
Density and phase of the order parameter corresponding to 
the critical frequency $\Omega_{c2}$ for the dipolar case, 
with $\sqrt{8 \pi} N a/a_z=150$, and $a_{dd}/a_0 = 2.505^2/3$.  
For (a) $\Theta = 0^{\circ}$, $\Omega_{c2}/\omega_0 \approx
0.80$, (b) $\Theta = 30^{\circ}$, $\Omega_{c2}/\omega_0 \approx
0.5835$, (c) $\Theta = 60^{\circ}$, $\Omega_{c2}/\omega_0 
\approx 0.36$, and (d) $\Theta = 90^{\circ}$, $\Omega_{c2}/
\omega_0 \approx 0.296$.}
\label{fig3}
\end{figure}

\section{Dipolar case}

We turn now to the case where the atoms have a nonzero
dipole moment. In addition to the hard-core potential 
considered in the previous section -- which is kept fixed, 
with $Mg/\hbar^2 = \sqrt{8 \pi} N a/a_z = 150$ -- we consider 
a dipolar interaction of a fixed strength. Introducing the 
dipolar length $a_{dd} \equiv M D^2/(3 \hbar^2)$, we choose 
$a_{dd}/a = 2.505^2/3$. In the above expression $D^2 = 
d^2/(4 \pi \epsilon_0)$ when the atoms have an electric dipole 
moment $d$, where $\epsilon_0$ is the permittivity of the 
vacuum; when the atoms have a magnetic moment $\mu$, $D^2 
= \mu_0 \mu^2/(4 \pi)$, where $\mu_0$ is the permeability 
of the vacuum. With this choice of parameters the dipolar 
and the contact interactions are of comparable magnitude. 

In this problem, small values of $D$ and thus weak 
dipolar interactions may give rise to a very smooth 
energy surface, which leads to degeneracy problems in 
the calculation. On the other hand, large values of $D$ 
tend to cure this problem, however they also make 
the system unstable against collapse for small values 
of the angle $\Theta$ due to the head-to-tail alignment 
of the dipoles.

We investigate the rotational response of the cloud considering 
four values for the angle of the dipole moment, $\Theta = 
0^{\circ}, 30^{\circ}, 60^{\circ}$, and 90$^{\circ}$, as in 
Fig.\,1. When $\Theta=90^{\circ}$ the interaction respects 
the axial symmetry of the trapping potential and is also 
purely repulsive. As a result, the system behaves 
qualitatively as in the case of a contact potential alone, 
with a phantom vortex state, a real vortex state, and a 
vortex lattice forming with increasing $\Omega$. 

For any other value of the angle $\Theta$, as $\Omega$ 
increases, initially the cloud still responds by forming 
one phantom vortex at the center of the trap, as in the 
non-dipolar case. On the other hand, for higher rotational
frequencies the gas behaves in a qualitatively different way. 
In this case the dipolar interaction breaks the axial symmetry 
of the Hamiltonian, introducing a preferred direction. 

In Fig.\,3 we plot the density and the phase of the order 
parameter for four values of the angle $\Theta = 0^{\circ}, 
30^{\circ}, 60^{\circ}$, and $90^{\circ}$, for the value of 
the angular frequency that corresponds to $\Omega_{c2}$, 
i.e., for the value of $\Omega$ for which the first (real) 
vortex state penetrates the annulus. While for $\Theta = 
90^{\circ}$ the position where the vortex enters the annulus 
is arbitrary (as in the non-dipolar case), for all other values 
of $\Theta$ this is determined by the direction of the 
polarizing field, which is chosen to be in the direction
going from the bottom to the top of the page. For $\Theta = 
60^{\circ}$ and $30^{\circ}$ there are actually two vortex 
states which enter the annulus from opposite ends. For 
$\Theta = 0^{\circ}$ instead of two vortices, there are 
two pairs of vortices, which are symmetrically displaced from 
the direction of the polarizing field. As one can see from the 
phase of the order parameter, as $\Theta$ decreases the number 
of phantom vortices increases and especially for $\Theta = 
0^{\circ}$ their number is very large. Also, with decreasing 
$\Theta$ the density at the center of the trap decreases. For 
$\Theta = 0^{\circ}$ the density in this region is extremely 
small and therefore it appears to be ``empty" in Fig.\,3,
as it can be seen also from the phase plot.

Furthermore, as we have seen numerically, the anisotropic 
character of the dipolar interaction affects the frequencies 
$\Omega_{c1}$ and $\Omega_{c2}$ in a different way. For 
$\Theta = 0^{\circ}, 30^{\circ}, 60^{\circ}$ and $90^{\circ}$, 
$\Omega_{c1}$ is found to increase linearly with $\Theta$, 
with $\Omega_{c1}/\omega_0$ being approximately equal to 0.035, 
0.036, 0.037, and 0.038, respectively. Also, $\Omega_{c2}/
\omega_0$ is found to be $\approx 0.80$, 0.5835, 0.36, and 
0.296, respectively. In other words, as $\Theta$ decreases, 
i.e., as the system gets more distorted from axial symmetry, 
$\Omega_{c1}$ decreases slightly (and linearly), whereas 
$\Omega_{c2}$ increases. It is interesting that for $\Theta
=30^{\circ}$, $\Omega_{c1}$ is identical to that of the 
non-dipolar gas. For this value of $\Theta$ the repulsive 
and attractive parts of the dipolar interaction are comparable 
and in a sense balance each other.
 
\begin{figure}[t]
\centerline{\includegraphics[height=8cm,width=9cm]{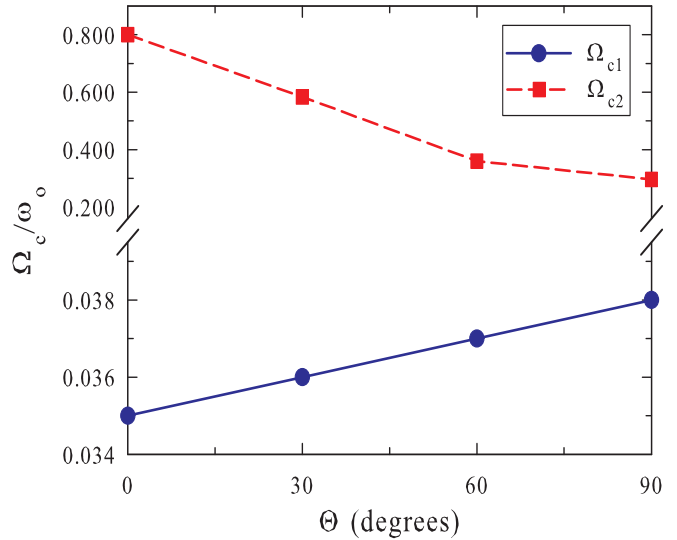}}
\caption{(Color online)
The critical frequencies $\Omega_{c2}$ and $\Omega_{c3}$
as a function of the angle $\Theta$ for the parameters of 
Fig.\,3.}
\label{fig4}
\end{figure}

\section{Summary}

In the present study we have investigated two problems. 
First, we studied the rotational response of a superfluid 
that is confined in an annular potential. Such a trapping 
geometry interpolates in a sense between one- and 
two-dimensional motion, depending on the ratio between 
the mean radius of the annulus and its width. The second 
problem we have addressed is the effect of the long-range 
and anisotropic character of the dipolar interaction.

Starting with the non-dipolar case, as the rotational
frequency of the annulus increases, there is a critical 
frequency above which a vortex state appears in the region 
of essentially zero density, around the center of the trap. 
Eventually more vortices start to reside in this region. 
These vortices affect only the phase of the order 
parameter, increasing the circulation and leaving the 
density unaffected. As the rotational frequency increases,
the number of ``phantom" vortices increases as well. 
Eventually the number of phantom vortices which can be 
accommodated in the region of low atom density reaches a 
threshold. As a result, above some critical frequency, one 
of these vortices penetrates the annulus, moving to the 
region of non-zero density. These ``real" vortices 
eventually form a regular vortex array, or even multiple 
arrays, with increasing $\Omega$.

The dipolar case differs from the above mainly because
of the symmetry-breaking of the Hamiltonian due to the 
dipolar interaction (unless the dipole moment is perpendicular 
to the plane of motion of the atoms). Even the non-rotating 
system is affected by the dipole-dipole interaction, with 
the formation of one or two density maxima, depending on the 
value of the parameters, as seen in Refs.\,\cite{Aba3, Aba10, 
MKR}. This inhomogeneity in the density affects also the 
critical rotational frequencies. Initially a ``phantom" 
vortex still appears at the center of the trap. However, 
for larger rotational frequencies of the trap the ``real" 
vortices penetrate the annulus either in pairs, or in 
couples of pairs from opposite ends of the cloud in the
direction of the polarizing field of the dipoles. 

Rotating atomic Bose-Einstein condensates confined 
in toroidal potentials have been realized experimentally 
recently, while dipolar gases are under intense experimental 
investigation. As shown in the present article, the combination 
of these two effects -- that should be accessible also 
experimentally -- gives rise to interesting effects.

\section{Acknowledgements}

We thank Alexander L. Fetter and Kimmo K\"arkk\"ainen for 
very useful discussions. This work was financed by the 
Swedish Research Council and originated from a collaboration 
within the POLATOM Research Networking Programme of the 
European Science Foundation (ESF). E. \"O. K. is supported 
by the Turkish Council of Higher Education (Y\"OK) within 
the scope of Post-Doctoral Research Scholarship Program.


\begin{thebibliography}{99}

\bibitem{rtr} S. Gupta, K. W. Murch, K. L. Moore, T. P. 
Purdy, and D. M. Stamper-Kurn, Phys. Rev. Lett. {\bf 95}, 
143201 (2005); S. E. Olson, M. L. Terraciano, M. Bashkansky, 
and F. K. Fatemi, Phys. Rev. A {\bf 76}, 061404(R) (2007);
K. Henderson, C. Ryu, C. MacCormick, and M. G. Boshier,
New J. Phys. {\bf 11}, 043030 (2009).

\bibitem{Ryu} C. Ryu, M. F. Andersen, P. Clad\'e, V. Natarajan, 
K. Helmerson, and W. D. Phillips, Phys. Rev. Lett. {\bf 99}, 
260401 (2007).

\bibitem{BPh} A. Ramanathan, K. C. Wright, S. R. Muniz, M. Zelan, 
W. T. Hill, C. J. Lobb, K. Helmerson, W. D. Phillips, and G. K. 
Campbell, Phys. Rev. Lett. {\bf 106}, 130401 (2011).
 
\bibitem{Foot} M. Gildemeister, B. E. Sherlock, and C. J. Foot
Phys. Rev. A {\bf 85}, 053401 (2012).

\bibitem{ZH} Stuart Moulder, Scott Beattie, Robert P. Smith, 
Naaman Tammuz, and Zoran Hadzibabic, eprint cond-mat/1112.0334.

\bibitem{Vief} S. Viefers, P. Koskinen, P. Singha Deo, and
M. Manninen, Physica E (Amsterdam) {\bf 21}, 1 (2004).

\bibitem{Mann} Manninen et al., to be published.

\bibitem{Grie} A. Griesmaier, J. Werner, S. Hensler, J. Stuhler, 
and T. Pfau, Phys. Rev. Lett. {\bf 94}, 160401 (2005);
J. Stuhler, A. Griesmaier, T. Koch, M. Fattori, T. Pfau,
S. Giovanazzi, P. Pedri, and  L. Santos, ibid.
{\bf 95}, 150406 (2005); M. Fattori, T. Koch, S. Goetz, 
A. Griesmaier, S. Hensler, J. Stuhler, and T. Pfau, Nature Phys.
{\bf 2}, 765 (2006); A. Griesmaier, J. Stuhler, T. Koch, 
M. Fattori, T. Pfau, and S. Giovanazzi, Phys. Rev. Lett.
{\bf 97}, 250402 (2006).

\bibitem{JYe} K. -K. Ni, et al., Science {\bf 322}, 231
(2008); S. Ospelkaus, K. -K. Ni, G. Quemener, B. Neyenhuis,
D. Wang, M. H. G. deMiranda, J. L. Bohn, J. Ye, and D. S. Jin,
Phys. Rev. Lett. {\bf 104}, 030402 (2010).

\bibitem{Lah} T. Lahaye, C. Menotti, L. Santos, M. Lewenstein, 
and T. Pfau, Rep. Prog. Phys. {\bf 72}, 126401 (2009).

\bibitem{Brunn} G. M. Brunn and E. Taylor, Phys. Rev. Lett.
{\bf 101}, 245301 (2008).

\bibitem{Aba3} M. Abad, M. Guilleumas, R. Mayol, M. Pi, and 
D. M. Jezek, Phys. Rev. A {\bf 79}, 063622 (2009). 

\bibitem{FMG11} F. Malet, T. Kristensen, S. M. Reimann, and
G. M. Kavoulakis, Phys. Rev. A {\bf 83}, 033628 (2011). 

\bibitem{Aba10} M. Abad, M. Guilleumas, R. Mayol, M. Pi, and 
D. M. Jezek, Phys. Rev. A {\bf 81}, 043619 (2010). 

\bibitem{MKR} F. Malet, G. M. Kavoulakis, and S. M. Reimann,
Phys. Rev. A {\bf 84}, 043626 (2011).

\bibitem{AB1} M. Abad, M. Guilleumas, R. Mayol, M. Pi, and 
D. M. Jezek, Phys. Rev. A {\bf 84}, 035601 (2011).

\bibitem{AB2} M. Abad, M. Guilleaumas, R. Mayol, M. Pi, and 
D. M. Jezek, Europhys. Lett. {\bf 94}, 10004 (2011).

\bibitem{AB3} M. Abad, M. Guilleaumas, R. Mayol, M. Pi, and 
D. M. Jezek, Laser Physics {\bf 18}, 648 (2008).

\bibitem{BLev} Mingwu Lu, Nathaniel Q. Burdick, Seo Ho Youn, 
and Benjamin L. Lev, Phys. Rev. Lett. {\bf 107}, 190401  (2011).

\bibitem{Zollner} Sascha Z\"ollner, G. M. Bruun, C. J. Pethick, 
and S. M. Reimann, Phys. Rev. Lett. {\bf 107}, 035301 (2011). 

\bibitem{Xio} B. Xiong, J. Gong, H. Pu, W. Bao, and B. Li,
Phys. Rev. A {\bf 79}, 013626 (2009); M. Asad-uz-Zaman and 
D. Blume, Phys. Rev. A {\bf 80}, 053622 (2009).

\bibitem{Smerzi} F. Piazza, L. A. Collins, and A. Smerzi,
Phys. Rev. A {\bf 80}, 021601 (2009). 

\bibitem{Pedri} R. Dubessy, T. Liennard, P. Pedri, and 
H. Perrin, Phys. Rev. A {\bf 86}, 011602 (2012).

\bibitem{Woo} S. J. Woo and Young-Woo Son, Phys. Rev. A 
{\bf 86}, 011604(R) (2012).

\bibitem{Fet67} Alexander L. Fetter, Phys. Rev. {\bf 153}, 
285 (1967).

\bibitem{Don67} P. J. Bendt and R. J. Donnelly, Phys. Rev. Lett.
{\bf 19}, 214 (1967).

\bibitem{Cre10} J. C. Cremon, G. M. Bruun, and S. M. Reimann, 
Phys. Rev. Lett. {\bf 105}, 255301 (2010).

\bibitem{Chi05} S. A. Chin and E. Krotscheck, Phys. Rev. E 
{\bf 72}, 036705 (2005).

\end{thebibliography}
\end{document}